\RequirePackage{fix-cm}
\documentclass[smallextended]{svjour3}       
\smartqed  
\usepackage{graphicx}
\usepackage{amsmath}
\usepackage{amssymb}
\usepackage{cite}
\usepackage{textcomp}
\usepackage{float}
\usepackage[dvipsnames]{xcolor}
\usepackage{soul}
\usepackage{ulem} 		
\journalname{ArXiv}
\begin{document}

\title{Noise performance \& thermalization of a single electron transistor using quantum fluids}
\titlerunning{Noise performance \& thermalization of SET using quantum fluids} 

\author{N.R.~Beysengulov \and J.R.~Lane \and J.M.~Kitzman \and K.~Nasyedkin \and D.G.~Rees \and J.~Pollanen}

\institute{N.R.~Beysengulov,  J.R.~Lane,  J.M.~Kitzman,  K.~Nasyedkin, J.~Pollanen \at
              Department of Physics and Astronomy, Michigan State University, East Lansing, Michigan 48824-2320, USA \\
              \email{beysengu@msu.edu}  
           \and
           D.G.~Rees \at
              EeroQ Corporation, Lansing, Michigan 48906, USA
}

\date{Received: date / Accepted: date}
\maketitle

\begin{abstract}
We report on low-temperature noise measurements of a single electron transistor (SET) immersed in superfluid $^4$He. The device acts as a charge sensitive electrometer able to detect the fluctuations of charged defects in close proximity to the SET. In particular, we measure telegraph switching of the electric current through the device originating from a strongly coupled individual two-level fluctuator. By embedding the device in a superfluid helium immersion cell we are able to systematically control the thermalizing environment surrounding the SET and investigate the effect of the superfluid on the SET noise performance. We find that the presence of superfluid $^4$He can strongly suppress the switching rate of the defect by cooling the surrounding phonon bath.
\keywords{single electron transistor \and liquid helium \and two-level fluctuator \and telegraph noise}
\end{abstract}

\section{Introduction}

Fluctuating charge traps or defects in amorphous materials are a main source of noise in solid state quantum devices. These fluctuations can originate from atomic-scale lattice defects that stochastically switch between two nearly equivalent configurations, and are usually called two-level fluctuators (TLFs)~\cite{phillips1987two}. A sparse bath of TLFs gives rise to ubiquitous low frequency $1/f$ noise, which is observed in all charge sensitive devices. Recently TLFs have attracted a renewed interest in the context of quantum information science due to the essential role of these defect in the coherence properties of quantum devices~\cite{Martinis2005,paladino20141,connors2019low,dial2013charge}. TLFs behave as quantum mechanical two-level systems that can couple to qubits via their electric dipole moments. Due to the broad frequency distribution of the splitting energies of TLFs, the qubit excitations can be transferred to TLFs leading to qubit depolarization~\cite{muller2015interacting,klimov2018fluctuations,schlor2019correlating,burnett2019decoherence}. Additionally, TLFs located directly within the tunnel junction of a superconducting qubit can cause critical current fluctuations leading to qubit dephasing~\cite{schlor2019correlating,burnett2019decoherence}.

Single electron transistors (SETs) are another class of charge sensitive quantum devices. When integrated into a high-frequency circuit containing an $LC$-resonator SETs are suitable for a wide variety of quantum measurements and these so-called RF-SETs have been used to investigate Cooper-pair boxes~\cite{bladh2005single,aassime2001radio}, quantum dots~\cite{fujisawa2000charge,yuan2012charge}, nanomechanical resonators~\cite{naik2006cooling} and single electron tunneling~\cite{bylander2005current,lu2003real}. At the heart of any SET device is a small conducting island, which is connected via tunnel junctions to source and drain electrodes (see Fig.~\ref{fig1}a). This island is capacitively sensitive to the electric field configuration of the surrounding environment. In particular, charge fluctuations due to thermal activation or quantum tunneling of a TLF located in close proximity \cite{krupenin1998noise,song1995advantages,henning1999bias,zorin1996background,wolf1997investigation,krupenin2000aluminium,brown2006electric,simkins2009thermal} leads to a noisy random telegraph signal in the electrical current flowing through the SET. 

Charge sensors based on SETs were proposed as a possible read-out device for a quantum computer architecture based on electrons trapped above the surface of superfluid helium~\cite{platzman1999quantum,lea2000could}. In these works the SET is submerged in the superfluid and located below an electron floating approximately 10~nm above the helium surface. A charge will be induced on the SET island depending on the charge state of the electron. In fact single electron trapping on the surface of liquid helium was demonstrated using a conventional SET~\cite{papageorgiou2005counting}. However to-date the noise performance of SET-based devices has not been investigated in the presence of liquid helium. Furthermore superfluid $^4$He is an ultra-clean dielectric devoid of defects and impurities. Integrating superfluid helium with other quantum systems, such as optomechanical~\cite{de2017ultra,harris2016laser}, micro-mechanical~\cite{Souris2017_2} or microwave~\cite{Souris2017,Clark2018} resonators and superconducting qubits~\cite{lane2020integrating}, provides a new tool for studying the quantum behavior and decoherence in these platforms.

In this work we study the noise properties of TLFs in the presence and absence of superfluid $^4$He by measuring the electrical transport through a conventional SET. We find that telegraph noise in the SET current originates from a strongly coupled TLF located in, or in close proximity, to an SET tunnel junction. The dependence of the characteristic TLF switching time on the SET bias voltage reveals thermally activated configurational changes of the TLF by hot electrons in the SET. At elevated temperatures we find a strong suppression of the TLF switching rate and a three-fold reduction of the TLF temperature due to the opening of an additional channel for the dissipation of heat into the liquid helium and the thermalisation of substrate phonons. At the lowest temperatures, in the presence of superfluid helium, the TLF is thermally decoupled from the substrate phonons, which we attribute to the poor thermal contact between electrons in the SET and phonons in the substrate.

\section{Experiments and results}
\begin{figure}
\centering
\includegraphics[width=0.75\textwidth]{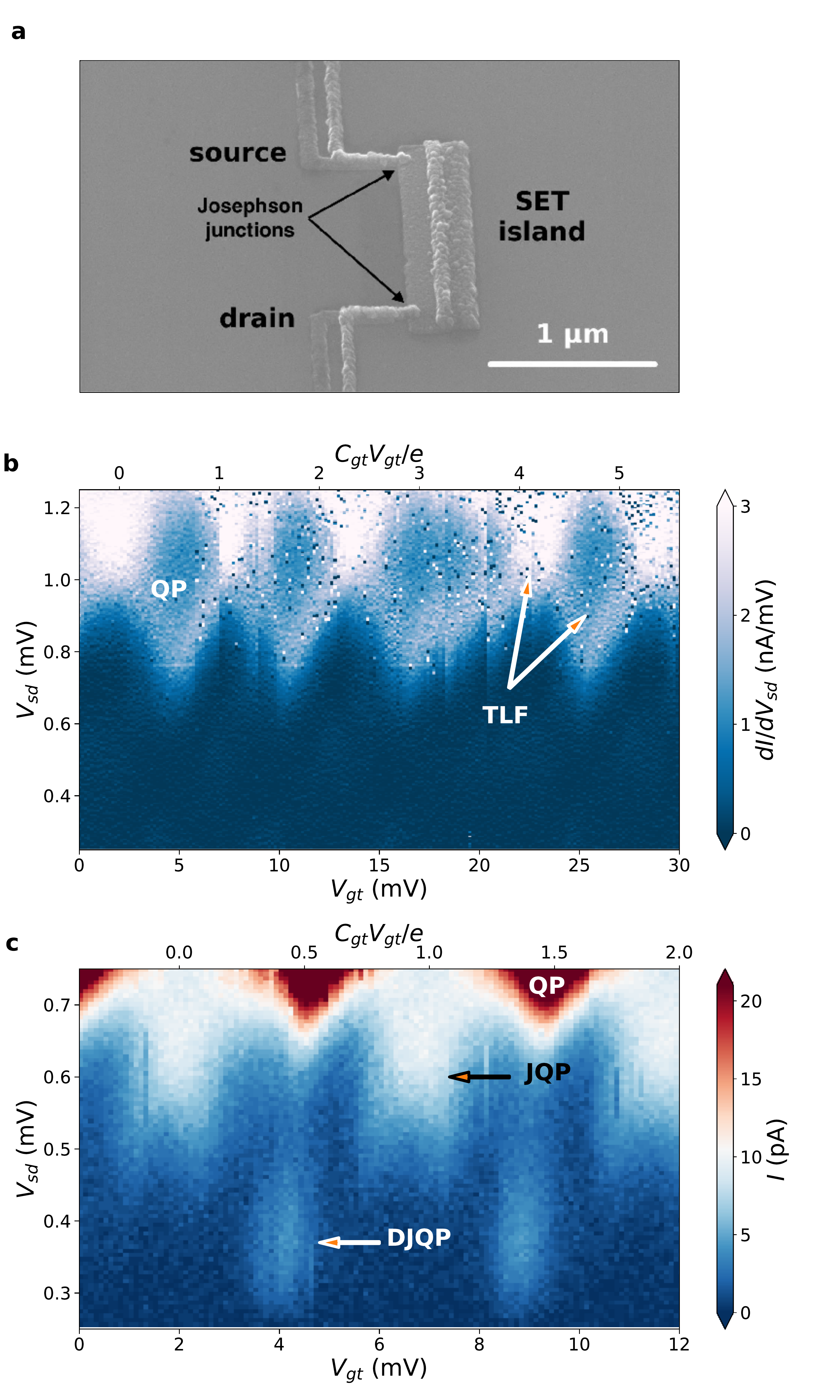}
\caption{\label{fig1} (a) Scanning electron microscope image of the superconducting SET device. The area of each tunnel junction is approximately 70~nm $\times$ 60~nm. The overlapping secondary rectangle on top of the SET island is formed by the two-angle (shadow) evaporation technique used to fabricate the tunnel junctions. A gate electrode is located outside the image, approximately 2~$\mu$m to the right of the island. (b) Derivative of the current through the SET as a function of $V_{\text{gt}}$ and $V_{\text{sd}}$ measured at 7~mK. (c) Measured current map of the SET in the superconducting gap region, which reveals various Cooper-pair tunneling processes (JQP and DJQP).}
\end{figure}

To examine the noise properties of SET-based devices we fabricated a conventional SET using two-angle evaporation of aluminum (Al) on silicon with a 500 nm thick oxide layer. The superconducting SET consists of two Al/AlO$_{\text{x}}$/Al tunnel junctions and an Al island having size of 1.5~$\mu$m $\times$ 0.3~$\mu$m $\times$ 0.02~$\mu$m (see Fig.~\ref{fig1}a). The source electrode of a second nearby SET, located 2~$\mu$m away, served as an external gate electrode allowing tuning of the induced charge on the SET island. The SET devices were mounted inside a copper superfluid leak tight sample box~\cite{nasyedkin2018unconventional} attached to the mixing chamber plate of a cryogen-free dilution refrigerator. The current through the SET was measured using a lock-in amplifier with the bias voltage $V_{\text{sd}}$ modulated at 36.9~Hz~\cite{rmsnote}.

The SET is characterized by a charging energy $E_{\text{C}} = e^2/2 C_{\sum}$, where $C_{\sum}$ is the total capacitance of the island, and the Josephson energy $E_{\text{J}} = \pi \hbar \Delta/(2 e)^2 R_i$, where $\Delta$ is the superconducting energy gap of Al and $R_i$~is the normal state resistance of a particular junction, $i=1,2$. There are two Josephson junctions and the measured total resistance of the device $R_T = R_1 + R_2 = 1.3$~M$\Omega$ leads to $\Delta \approx E_{\text{C}} \gg E_{\text{J}}$, a condition which promotes Cooper-pair tunneling processes~\cite{hadley19983}. Figure~\ref{fig1}b shows the derivative of the measured current through the superconducting SET at different bias and gate voltages. Coulomb diamond structures emerge at bias voltages $V_{\text{sd}} > 0.65$~mV, which are formed from the threshold for sequential quasiparticle (QP) tunneling through the SET. These tunneling processes require a minimum bias voltage of $4 \Delta /e$ and maximum $4 \Delta /e + e/C_{\sum}$. $C_{\sum} = C_{\text{gt}} + C_1 + C_2$ is the total capacitance, $C_{\text{gt}}$ is the capacitive coupling between the gate and the island, $C_1$ and $C_2$ are the capacitances of the Josephson junctions. The slope of the threshold lines in Fig.1b and Fig.1c are given by $-C_{\text{gt}}/C_2$ and $C_{\text{gt}}/(C_1 + C_{gt})$. At low bias voltages the tunneling of individual quasiparticles is suppressed due to a combination of Coulomb blockade and the absence of states in the superconducting gap. We also note abrupt shifts in the Coulomb diamond structure as a function of $V_{\text{gt}}$, which we attribute to random changes in the induced electric field from background charge traps. The current measurements inside the superconducting gap region, shown in Fig.~\ref{fig1}c, reveal other tunneling processes, known as Josephson-quasiparticle (JQP) and double Josephson-quasiparticle (DJQP) cycles~\cite{hadley19983}. JQP cycles, seen as intersecting ridges centered at $V_{\text{sd}} \approx 0.62$~mV in Fig.1c, occur when a Cooper pair is transported through one of the SET junctions while two quasiparticles are transported through the other junction. The bias voltage at the JQP current intersections is given by $2 e/C_{\sum}$. The isolated current peaks located at a bias of approximately 0.35~mV correspond to DJQP cycles (usually observed at a bias voltage of $e/C_{\sum}$), where two resonant Cooper pair tunneling events (with different parity) and two quasiparticle events are involved. During the DJQP cycle, $3e$ of charge is transported through SET. The SET parameters extracted from these data are summarized in a Table~\ref{tab:tab1}.

\begin{table}
\caption{\label{tab:tab1}SET parameters.}
\begin{tabular}{c c c c c c c}
\hline\noalign{\smallskip} 
 $R_T$ &$\Delta$ &$E_C$ &$C_1$ 
 &$C_2$ &$C_{\text{gt}}$ \\ [0.5ex] 
\noalign{\smallskip}\hline\noalign{\smallskip}
 1.3 M$\Omega$ & 180 $\mu$eV& 165 $\mu$eV & 0.24 fF & 0.19 fF
& 0.03 fF \\
\noalign{\smallskip}\hline 
\end{tabular}
\end{table}

\begin{figure}
\centering
\includegraphics[width=0.7\textwidth]{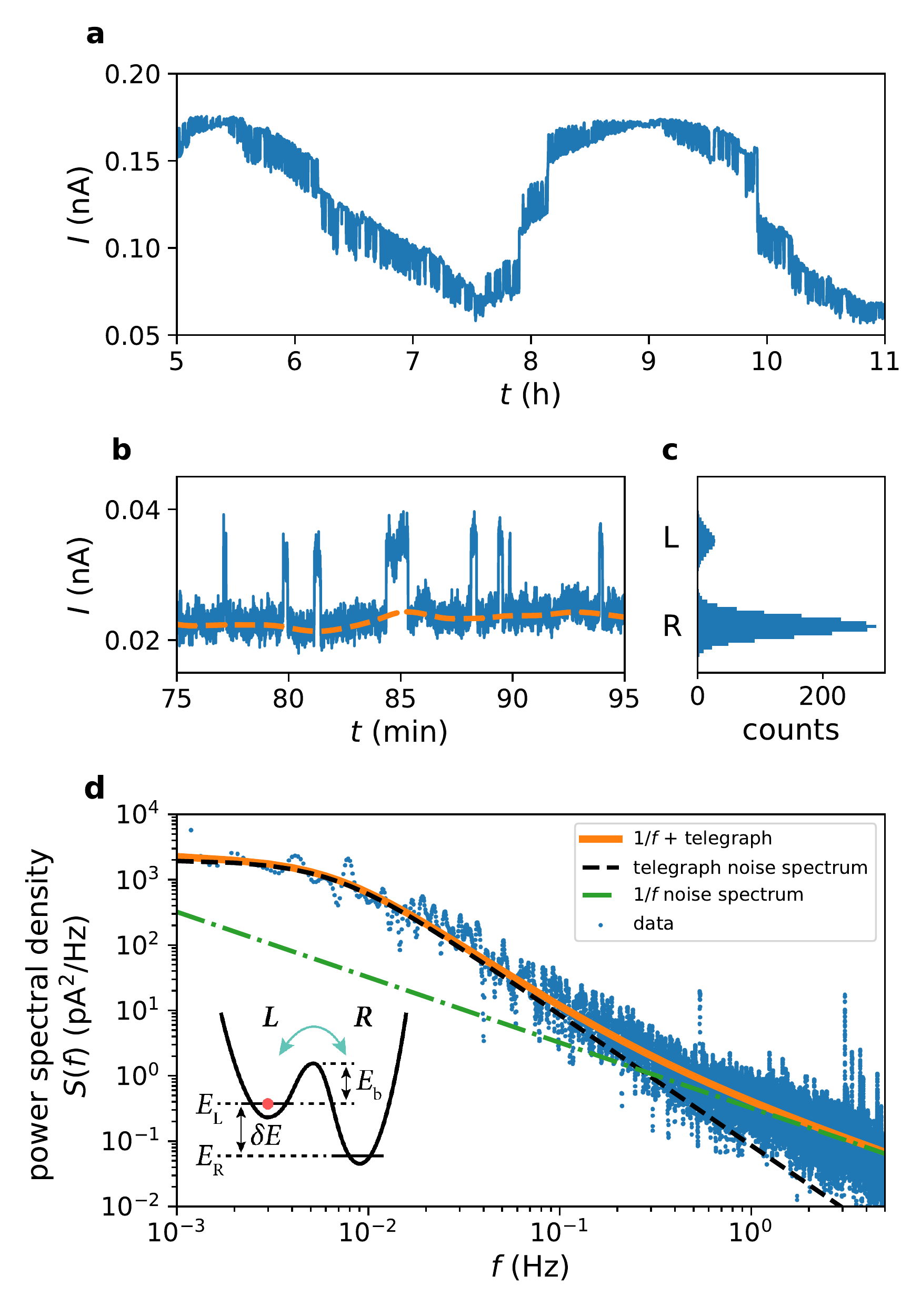}
\caption{\label{fig2} (a) Measured SET current showing three type of noise with different time scales. (b) The current through the SET measured at $V_{\text{sd}} = 0.85$~mV as a function of time demonstrates random telegraph noise. The additional small drift of the current is tracked using an asymmetric least squares method (dashed orange line). (c) SET current histogram corresponding to the occupation of a strongly coupled TLF. (d) Power spectral density of the SET current fluctuations. The contribution of Lorentzian and $1/f$ spectra to the overall fit (solid orange line) are given by dashed (black) and dashed-dotted (green) lines, respectively. The inset shows the energy model of TLF.}
\end{figure}

Time-resolved traces of the current $I(t)$ through the SET provide valuable information about charge noise sources present in the system. Figure~\ref{fig2}a shows an example of a current trace acquired at $T$ = 7~mK and $V_{\text{sd}} = 0.85$~mV. Here the current was sampled continuously after the gate voltage was stepped from $V_{\text{gt}} = 5.3$~V to $V_{\text{gt}} = 0$~V. The long-scale current drift originates from the change in the potential landscape of the TLFs located in the region between gate electrode and the SET island~\cite{black-halperin1977}. According to the model developed in ref.~\cite{black-halperin1977} abrupt change in the gate voltage causes a change in the potential differences between two wells of these TLFs. Thus some of the TLFs can be brought to metastable states that subsequently decay to lower energy charge states. A new equilibrium charge distribution is reached after some characteristic time causing a drift of the induced charge on the SET island. The second type of noise observed in $I(t)$ is a cascade of signal jumps on the time scale of several hours. These current jumps can be attributed to more strongly coupled TLF clusters~\cite{zimmerman1997modulation}. Noise from these clusters is also responsible for the random shifts along the horizontal axis on the differential conduction map as shown in Fig.~\ref{fig1}b.

The most prominent noise source observed is the switching of the SET current between two states on a time scale of several minutes. This type of noise is characteristic of a single TLF strongly coupled to the SET. The magnitude of the current jumps is modulated as the background charge environment drifts in time and seems to vanish at $t = 9$~h in Fig.\ref{fig2}a. This behavior is consistent with the TLF creating an induced charge modulation $\delta q \approx 0.06$~$e$ on the SET island rather than causing a change in the transparency of either of the tunnel junctions. We note that the switching rate of the TLF does not depend on the time varying background charges. Thus, we can conclude that the two-level noise source is located inside or in close proximity to one of the the tunnel junctions, where it is shielded by the Al leads and island making it insensitive to external electric fields. We also note that this telegraph noise appears mostly in the QP tunneling transport regime and is rarely observed inside the superconducting gap region indicating thermal activation of the switching process. The small drift of the current in Fig.~\ref{fig2}b (orange dashed curve) possibly originates also from the large bath of weakly coupled TLFs with distinct coupling and telegraph jumping rates. This type of noise has been observed in different superconducting qubit platforms and is responsible for a continuous drift of the resonant frequency and decoherence rate of the qubit, which is usually discussed in terms of spectral diffusion~\cite{klimov2018fluctuations}. Here we use an asymmetric least square method to subtract this drift in order to perform a proper statistical analysis of current traces.

In a simple model the microscopic configuration of a TLF corresponds to a charge particle trapped in a double-well potential having an energy difference $\delta E = E_{\text{L}} - E_{\text{R}}$ between the left and right wells which are separated by a potential barrier $E_{\text{b}}$ (see inset of Fig.~\ref{fig2}d). Thermally activated stochastic motion of a charge in this double-well trap is characterized by the dwell times $\tau_{L}$ and $\tau_{R}$, which depend on the properties of the TLF potential and the TLF temperature $T_{\text{TLF}}$. In thermal equilibrium the ratio of the dwell times can be calculated using Boltzmann statistics $\tau_{L}/\tau_{R} = \text{exp} ( - \delta E/k_{\text{B}} T_{\text{TLF}} )$; where $k_{\text{B}}$ is the Boltzmann constant. The histogram of the population probabilities of the TLF states extracted from the measured current traces provides direct information about the ratio $\tau_{L}/\tau_{R}$. Such a histogram, shown in Fig.~\ref{fig2}c, demonstrates the highly asymmetric structure of the TLF double-well. The random telegraph noise signal produces a power spectral density (PSD) having a Lorentzian form $S(\omega) \propto \bar{\tau}/(1 + \omega^2 \bar{\tau}^2)$ centered at zero frequency. Here $\omega = 2 \pi f$ and $\bar{\tau} = 1/\gamma$ is the TLF switching time, where $\gamma$ is the sum of forward $1/\tau_{R}$ and backward $1/\tau_{L}$ switching rates. Note, that typical low-frequency noise of the form $\sim 1/f^{\alpha}$, usually observed in many solid state quantum devices, originates from a large ensemble of two-level defect with a superposition of many such Lorentzian spectra. The PSD, shown in Fig.~\ref{fig2}d, is reasonably described by a single Lorentzian added to a $1/f$-type background \cite{verbrugh1995optimization,kenyon2000temperature,kafanov2008charge}. For $V_{\text{sd}} = 0.85$~mV we obtain a TLF switching time $\bar{\tau} = 150$~s from a fit to the theoretical PSD and an energy difference $\delta E/k_{\text{B}} T_{\text{TLF}} = 2.1$, which is extracted from the ratio of the population probabilities of the TLF’s states. The standard deviation of current drift evolves in time diffusively as $\sigma(t) = 2 D t^{1/2}$ with the diffusivity $D = 0.7$~pA~(hour)$^{-1/2}$. We estimate the upper limit for the charge sensitivity of our SET device to be $2 \times 10^{-3} \: e/\sqrt{\text{Hz}}$ at 1~Hz, which is estimated from the current PSD and the current-charge transfer function of the SET. 

\begin{figure}
\centering
\includegraphics[width=1.0\textwidth]{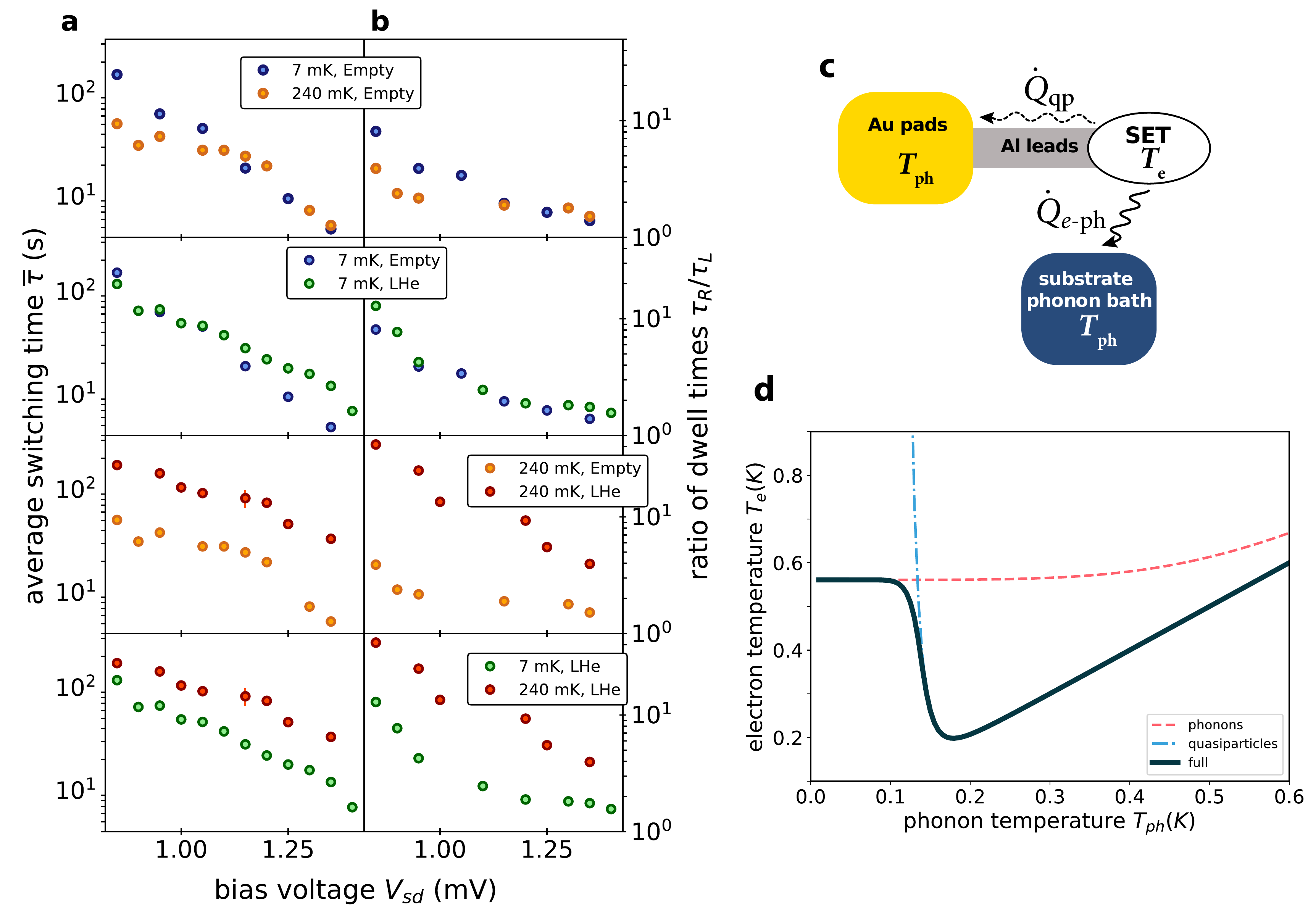}
\caption{\label{fig3} (a) The measured average switching times extracted from the noise PSD and (b) the ratio of dwell times of the strongly-coupled TLF extracted from occupation histograms at two different temperatures in the presence and absence of liquid helium in the sample cell. (c) The heat dissipation model of the SET device (described in detail in the main text) (d) The electron system temperature extracted from the heat balance equation $P_{\text{SET}} = \dot{Q}_{\text{qp}} + \dot{Q}_{e-\text{ph}}$ with $P_{\text{SET}} = 0.2$~pW (solid line). Dashed and dot-dashed lines show the contributions from direct relaxation to phonons and heat transfer through the superconducting leads, respectfully.}
\end{figure}

In order to understand the thermal properties of the TLF we measured the switching time and the ratio of dwell times at different bias voltages, temperatures and by controlling the thermalizing environment surrounding the SET via the introduction of liquid $^4$He (see Fig.~\ref{fig3}a,b). Our results indicate that the TLF temperature is lowest when the cryostat temperature is $T_{\text{mxc}} = 240$~mK and the experimental cell is full of liquid helium, while measurements under different conditions (either at lower cryostat temperature or in the absence of liquid helium) indicate a higher defect temperature. 

To understand these experimental observations, we start our discussion with an analysis of the quantities $\bar{\tau}$ and $\tau_L/\tau_R$ in the absence of liquid helium surrounding the SET (as shown in the top panels of Fig.~3a,b). These measurements show that both quantities decrease with increasing bias voltage, indicating a thermally activated switching process. At high bias voltages $V_{sd} > 1.1$~mV there is almost no difference between the quantities $\bar{\tau}$ and $\tau_L/\tau_R$ measured at different cryostat temperatures indicating that the defect temperature $T_{\text{TLF}}$ is determined predominantly by the applied power. 

In general, the TLF can be activated either by the inelastic scattering of hot electrons in the SET, in which case we expect $T_{\text{TLF}}$ to follow the electron temperature $T_e$, or by the local heating of the substrate in the proximity of the SET by the hot electrons, in which case the TLF temperature should follow the local phonon temperature $T_{\text{ph}}$. We note that quasiparticles near the tunnel junction rapidly relax to a Fermi-Dirac distribution due to fast electron-electron interactions, and that the energy is therefore dissipated to the electron system with temperature $T_e$. The electron-phonon coupling provides a heat dissipation channel into the substrate phonon bath with temperature $T_{\text{ph}}$ and can become a bottleneck for heat flow at low temperatures, so that $T_e > T_{\text{ph}}$. However, it has been shown that under certain conditions the heat flow from overheated electrons in SETs into the substrate can locally increase the substrate temperature close to the SET~\cite{savin2006thermal}. To investigate the temperature distribution in the substrate we performed finite element modelling of the heat flow in our device, following ref.~\cite{gustafsson2013thermal}. We used temperature dependent thermal conductivities $\kappa_{\text{Si}} = 5.0 T^3$/K$^{3}$WK$^{-1}$m$^{-1}$ for Si and $\kappa_{\text{SiO}_{2}} = 0.03 T^2$/K$^{2}$WK$^{-1}$m$^{-1}$ for the 500~nm thick SiO$_2$~\cite{gustafsson2013thermal}. In our modeling the total dissipated power $P_{\text{SET}} = V_{\text{sd}} I$ is provided by a heat source having the actual size of the SET at the Al/SiO$_2$ interface, and the substrate temperature $T_{\text{ph}}$ at the boundaries is varied. We find that for $T_{\text{ph}} = 10$ mK the substrate temperature near the SET reaches $80$~mK for the highest dissipated power $P_{\text{SET}} = 0.5$~pW. For $T_{\text{ph}} > 200$~mK the dissipated power results in a negligibly small increase in temperature $\Delta T_{\text{ph}} < 5 \%$ near the SET location. This is in contrast to our experimentally obtained ratio of TLF temperatures for the lowest and highest dissipated powers  $T_{\text{TLF}}^{\text{high}P}/T_{\text{TLF}}^{\text{low}P} \simeq 2$. Therefore, the local heating of the substrate in the vicinity of the SET alone cannot completely explain our experimental observations shown in the top panel of Fig.~3a,b~\cite{rmsnote2}. 

For the case in which the TLF is activated directly by electrons in the SET, we assume that the TLF temperature should follow the electron temperature. We believe there are two main energy relaxation channels for electrons in the SET (see Fig.~\ref{fig3}c). First is the coupling to phonons, which is described by $\dot{Q}_{e\text{-ph}} = \Sigma \Omega (T_e^n - T_{\text{ph}}^n)$ with $\Sigma = 0.4 \times 10^9$~W~K$^{-5}$~m$^{-3}$ being a material constant for Al, $\Omega$ is the volume of the SET and $n = 5$~\cite{wellstood1994hot}. The second relaxation channel is provided via thermal quasiparticles present in the superconducting leads, the density of which are suppressed exponentially at low temperatures. To qualitatively describe this mechanism we make several assumptions. First, we note that one end of the superconducting leads is connected to Au electrodes of large surface area and we have calculated that in this area the electron temperature will follow the substrate phonon temperature down to 50~mK. Second, we assume the power transferred by quasiparticles through the superconducting SET leads is given by $\dot{Q}_{\text{qp}} = 2\kappa_{\text{Al}} \nabla T = 2\kappa_{\text{Al}}(T) (T_e - T_{\text{ph}})S/L$, where $\kappa_{\text{Al}}(T) = \kappa_s e^{-\beta T_c/T_{\text{ph}}}$ with $\kappa_s$ = 120~W~K$^{-1}$~m$^{-1}$ and $\beta = 1.36$ for Al~\cite{satterthwaite1962thermal}, $T_c$ is the superconducting transition temperature, $S = 100\times40$~nm$^2$ and $L = 10$~$\mu$m are the cross-section of the  superconducting leads and the length, respectfully. With this simple model we numerically solve the heat balance equation $P_{\text{SET}} = \dot{Q}_{\text{qp}} + \dot{Q}_{e-\text{ph}}$ for the electron temperature $T_e$ at different substrate phonon temperatures, the results of which are shown as the solid black curve in Fig.~\ref{fig3}d. Additionally, Fig.~\ref{fig3}d also presents the independent contributions from quasiparticles and phonons shown as the blue dot-dashed and red dashed curves respectively. At high temperatures the heat transfer through the superconducting leads is the dominant mechanism, and electrons in the SET remain well thermalised to the phonon bath. However, at low temperatures $T_{\text{ph}} < 200$ mK this mechanism is strongly suppressed due to the extremely low quasiparticle density. In this case, heat can dissipate only through the electron-phonon coupling channel which, at these temperatures, provides a bottleneck that results in the overheating of the electron system. However, our experimental observations, which provide a measure of the TLF temperature, do not show this behaviour. This implies that, in the absence of liquid helium, the substrate phonon temperature is higher than the cryostat temperature, likely due to a relatively weak thermal connection between the substrate and the cryostat mixing chamber plate.

Introducing liquid helium into the sample cell provides an additional channel for heat dissipation in the system. (We also note that, in the presence of liquid helium with dielectric permittivity $\varepsilon = 1.056$, the capacitance between the gate electrode and the island increases by 2.2\%.) The liquid helium is introduced into the sample cell through a volume containing silver sinter, which provides a large surface area in thermal equilibrium with the cryostat mixing chamber and enhances the thermal coupling to the liquid. 
At the elevated cryostat temperature of 240 mK, the increase in the TLF switching time associated with the introduction of liquid helium is clear (see the third panels of Figure~\ref{fig3}a,b). We estimate the ratio of the defect temperature with and without helium $T_{\text{TLF}}^{\text{He}}/T_{\text{TLF}}^{\text{Empty}} = \log(r^{\text{Empty}})/\log(r^{\text{He}}) \simeq 1/3$, where $r = \tau_R/\tau_L$. This can be explained by the additional cooling of the substrate by the liquid helium, which assists in the thermalization of substrate phonons to the cryostat temperature. 
The heat balance within the device renormalizes in the presence of liquid helium, which results in a more effective cooling of the substrate and the TLF. At low temperatures the heat transfer through a solid-liquid interface is governed by the Kapitza boundary resistance which arises due to the acoustic mismatch between superfluid helium and a solid. In general, the Kapitza boundary resistance $R_K \propto T^{-i}$ with $i$ ranging between 1 and 3, which makes heat transfer at low temperatures extremely difficult~\cite{pobell2007matter,ramiere2016thermal}. Near the base cryostat temperature we see an increase in the TLF temperature for all values of the excitation power, which can be understood through the model presented in the Fig.~\ref{fig3}c, where $T_{\text{ph}} = T_{\text{mxc}}$. We note that at the lowest cryostat temperature (7~mK) we do not observe any difference in the TLF switching time parameters with or without helium (see Fig.~\ref{fig3}b). This indicates a relatively high TLF temperatures ($\sim 0.7$ K) and consequentially high electron and phonon temperatures in the absence of liquid helium in the cell, which can be explained by the relatively weaker thermalization of the substrate to the mixing chamber plate in this configuration.

The model we have presented here is based on several assumptions, which allow us to relate the key heat dissipation processes to the experimental observations. However, the true nature of the heat dissipation could be more complex. For example, the microscopic mechanism of the TLF noise usually assumes the movement of charges between different localized states in the tunnel barrier, or the trapping of electrons in Kondo-like subgap states localized near the superconducting-insulator boundary~\cite{faoro2006quantum,faoro2007microscopic}. Although we cannot definitively rule out these effects, we can argue that the TLF does not change the transparency of the tunnel junction of the SET. 
Regardless, in the context of the relatively simple model presented here, the presence or absence of liquid helium at elevated temperatures significantly renormalizes the heat balance equation, directly affecting the properties of the TLF noise source coupled to the SET. In this fashion the TLF can act as a local probe providing a measure of the temperature of electrons in the SET or the local phonon temperature. Controllable local thermometry can be utilized in interface heat transport studies~\cite{giri2020review} and can aid the understanding of heat flow in micro-/nano-electronic devices~\cite{jones2020progress}. To achieve a good thermal contact between liquid helium and solids at low temperatures one can use liquid $^3$He~\cite{Samkharadze2011}, which has a significantly lower thermal boundary resistance~\cite{pollanen2009low}. This provides a potential route to cooling the phonon bath in the substrate and achieving defect temperatures below 10 mK.

\section{Conclusion}
In conclusion, we have measured the noise performance of a single electron transistor in the presence and the absence of liquid helium. The transport properties of the SET are strongly affected by an individual two-level fluctuator located inside or in close proximity to a SET tunnel junction. The thermal properties of the TLF, which are embodied in its state-switching processes, can be significantly modified by the presence of liquid helium, which provides an extra cooling channel for the phonons in the substrate, reducing the frequency of switching events of the TLF.

\begin{acknowledgements}
We thank M.I. Dykman, N.O. Birge, H. Byeon, L. Zhang, C. Mikolas, and B. Arnold for fruitful discussions. We also thank R. Loloee and B. Bi for technical support and use of the W.M. Keck Microfabrication Facility at MSU. This work was supported by a sponsored research grant from EeroQ Corp. J. Pollanen and D.G. Rees are co-founders and partial owners of EeroQ Corp. Additionally, J. Pollanen, J.R. Lane and J.M. Kitzman acknowledge support from the National Science Foundation via grant numbers DMR-170833 and DMR-2003815 as well as the valuable support of the Cowen Family Endowment at MSU.

The data that support the findings of this study are available from the corresponding author upon reasonable request.
\end{acknowledgements}

\bibliographystyle{spphys}       

\begin{thebibliography}{10}
\bibitem{rmsnote}
The measured current and $V_{\text{sd}}$ are presented in rms units.
\end{thebibliography}

%
%

\end{document}